\documentclass[aps,prc,superscriptaddress,twoside,twocolumn,showpacs]{revtex4-1}
\usepackage{amsmath,amssymb}
\usepackage{graphicx}
\begin{document}

\title{Measurement and analysis of the $\boldsymbol{pp\to pp\gamma}$ reaction at
310~MeV}

\author{A.~Johansson}\email{ritaj@telia.com}%
\affiliation{Department of Physics and Astronomy, Uppsala
University, Box 535, S-751 21 Uppsala, Sweden}%
\author{H.~Haberzettl}\email{helmut@gwu.edu}
\affiliation{Center for Nuclear Studies, Department of Physics, The
George Washington University, Washington, DC 20052, U.S.A.}
\author{K.~Nakayama}\email{nakayama@uga.edu}%
\affiliation{Department of Physics and Astronomy, University of
Georgia, Athens, GA 30602, U.S.A.}%
\affiliation{\mbox{Institut f{\"u}r Kernphysik and J\"ulich Center
for Hadron Physics, Forschungszentrum J{\"u}lich, D-52425 J{\"u}lich,
Germany}}%
\author{C.~Wilkin}\email{cw@hep.ucl.ac.uk}%
\affiliation{Physics and Astronomy Department, UCL, London WC1E 6BT,
United Kingdom}%

\date{\today}

\begin{abstract}
The $pp\to pp\gamma$ reaction has been studied at a beam energy of
310~MeV by detecting both final protons at the PROMICE-WASA facility
and identifying the photon through the resulting missing-mass peak.
The photon angular distribution in the center-of-mass system and
those of the proton-proton relative momentum with respect to the beam
direction and to that of the recoil photon were determined reliably
up to a final $pp$ excitation energy of $E_{pp}\sim 30$~MeV. Except
for very small $E_{pp}$ values, the behavior of these distributions
with excitation energy is well reproduced by a new refined model of
the hard bremsstrahlung process. The model reproduces absolutely the
total cross section and its energy dependence to within the
experimental and theoretical uncertainties.
\end{abstract}


\maketitle

%
%

\section{Introduction}
\label{sec:intro}

The classic motivation for measuring the emission of bremsstrahlung
in nucleon-nucleon collisions was the study of the off-shell behavior
of the associated elastic scattering amplitude, although it is now
known that off-shell effects cannot be measured, even in principle.
Nevertheless, the bremsstrahlung reaction can provide a window into
the underlying dynamical mechanisms that govern the $NN$ interaction
and the electromagnetic currents of nucleons and mesons alike. This
is especially true for the hard part of the bremsstrahlung spectrum,
where the photon takes a large fraction of the available
center-of-mass energy. In this region the $\Delta(1232)$ isobar may
also play some role and the reaction has then much in common with
meson production.

Hard bremsstrahlung has been studied in the radiative capture in
neutron-proton scattering, $np\to d\gamma$, to quite high energies
through the measurement of the inverse reaction of deuteron
photodisintegration~\cite{CRA1996,ROS2005}. The energy dependence of
the cross section provides direct evidence that one of the main
driving terms is the excitation of an $S$-wave $\Delta N$ pair that
de-excites through an $M1$ transition into $d\gamma$.

The situation is radically different in proton-proton collisions. The
analogous $M1$ transition is forbidden for $pp\to pp\gamma$ by
angular momentum and parity conservation when the two protons emerge
with very low excitation energy $E_{pp}=W_{pp}-2m_p$, where $W_{pp}$
is the total energy of the final $pp$ subsystem in its rest
frame~\cite{LAG1989}. There are therefore significant cancelations
among the large amplitudes in the pure $S$-wave diproton limit, so
that any $\Delta$ effect must enter in more subtle ways. Furthermore,
the $E1$ transition is generally suppressed by the vanishing of an
electric dipole operator for the proton pair. It is therefore to be
expected that the bremsstrahlung production rate should be much lower
in $pp$ collisions than in $np$.

One technique used to investigate the emission of hard bremsstrahlung
in proton-proton collisions is the photodisintegration of a $pp$ pair
in $^3$He. Events where two fast protons emerge from
$\gamma^{3}\textrm{He}\to ppn$ are interpreted in terms of an
interaction on a diproton, with the neutron merely appearing as a
\textit{spectator}~\cite{He3,EMU1994,TED1994}. Such data show little
evidence for the presence of an intermediate $\Delta N$ pair,
certainly much less than for those with fast $pn$
pairs~\cite{EMU1994,TED1994}. However, because the capture on $pn$
spin-triplet pairs is so much stronger, the $pp$ data extracted in
this way may be contaminated by final state interactions, possibly
involving $np$ charge exchange. This can only be checked through
direct $pp\to pp\gamma$ measurements.

Proton-proton bremsstrahlung has been studied in several experiments
but, in general, these were undertaken by detecting the emerging
protons in pairs of small counters, often placed on either side of
the beam direction~\cite{pp_brems,MIC1990}, which has led to the low
$E_{pp}$ region being especially poorly sampled. The geometric
acceptance was much increased in a series of refined KVI experiments
at 190~MeV~\cite{KVI02,KVInn}, but even here the low $E_{pp}$ region
was not favored. Whereas the COSY time-of-flight spectrometer also
has wider acceptance, the data obtained at 293~MeV have only limited
statistics and no attempt was made to evaluate the cross section as a
function of $E_{pp}$~\cite{BIL1998}.

Data on the hardest part of the $pp$ bremsstrahlung spectrum were
also obtained at the COSY-ANKE magnetic spectrometer by selecting the
two final protons with $E_{pp}<3$~MeV~\cite{KOM2008,TSI2010}. A
proton beam energy range from 353 to 800~MeV was investigated but
only for CM photon angles $\theta_{\gamma}$, where
$\cos\theta_{\gamma}>0.95$. The results reveal a broad peak in the
cross section at an energy around 650~MeV with a FWHM~$\approx
220$~MeV. This suggests the possible influence of intermediate
$\Delta N$ pairs, though not necessarily in a relative $S$-wave.

Much higher statistics were obtained over a wider range of $pp$
excitation energies and photon angles at the PROMICE-WASA facility at
Uppsala. The experiment was carried out at a single beam energy of
$T_p=310$~MeV and results were recently published for
$E_{pp}<3$~MeV~\cite{JOH2009}. These data are completely consistent
with those from ANKE at 353~MeV over the small-angle domain covered
by the ANKE experiment~\cite{KOM2008,TSI2010}. However, it is clear
from this comparison that a reliable decomposition into multipoles
requires data over a wide angular range. The low $E_{pp}$ data from
Uppsala were interpreted as indicating the dominance of the $E1$ and
$M2$ multipoles~\cite{JOH2009} with no evidence for any important
$E2$ contribution, in contrast to theoretical
expectations~\cite{WIL1995}. The purpose of the present paper is to
extend the analysis up to $E_{pp}\approx 30$~MeV in order to test
theoretical models over a wider range of excess energies.

A state-of-the-art model has recently been developed that for the
first time describes successfully proton-proton bremsstrahlung in the
hundred MeV range~\cite{HN10,NH09}. This model, which is hereinafter
denoted as HN, is summarized in Sec.~\ref{sec:model}. In this
approach the photon is coupled everywhere to a relativistic $pp$
scattering amplitude in a way that ensures consistency with gauge
invariance. Although this reproduces very well the detailed KVI
$pp\to pp\gamma$ measurements at 190~MeV~\cite{KVI02}, it is possible
that at 310~MeV the tail of the $\Delta$ might have some influence.
In this context it should be noted that the minimal inclusion of the
$\Delta$ isobar~\cite{dJNL95} (see also Ref.~\cite{HN10}) improves
the theoretical description of the 280~MeV TRIUMF
data~\cite{MIC1990}.

The experimental approach used in this work is identical to that
employed at PROMICE-WASA for pion production~\cite{ZLO98,BIL01} and
so Sec.~\ref{sec:experiment} and Sec.~\ref{sec:analysis} merely
provide outlines of the salient points of the method and the data
analysis, respectively. The results given in Sec.~\ref{sec:results}
show that, away from the region of small $E_{pp}$ values, where there
can be significant cancelations between different contributions, the
theory of Sec.~\ref{sec:model} works remarkably well. It describes
the photon angular distribution and those of the diproton relative
momentum in different $E_{pp}$ intervals as well as the energy
dependence of the $pp\to pp\gamma$ total cross section. The fact that
the theory reproduces the absolute normalization of these
high-momentum-transfer data to within the experimental and
theoretical uncertainties is striking. However, the theoretical
predictions of the photon angular distributions obtained without
intermediate $\Delta N$ contributions are better at low $E_{pp}$ than
those that include them. This brings into question whether the
present simplified treatment of these isobar contributions is
acceptable. Our conclusions and suggestions for further work are to
be found in Sec.~\ref{sec:conclusions}.

%
%
\section{Theoretical model}
\label{sec:model}

The experimental data presented in this paper represent the most
complete measurement of bremsstrahlung in proton-proton collisions at
an energy so far above threshold. We therefore compare them with a
refined model~\cite{HN10} that has recently been successfully
applied~\cite{HN10,NH09} to describe both the TRIUMF data at
280~MeV~\cite{MIC1990} and the high-precision KVI data~\cite{KVI02}
at the lower proton energy of $T_p=190$~MeV. This solved a
long-standing discrepancy between experiment and the, then, existing
theory. The novel approach is derived within a quantum field-theory
formalism by coupling the photon everywhere possible to an underlying
two-nucleon $T$-matrix that is derived from a relativistic $NN$
scattering equation. The basic idea of the method is that introduced
by Haberzettl, Nakayama, and Krewald~\cite{HNK06} for pion
photoproduction, based on the field-theoretical approach of
Haberzettl~\cite{H97}. The model accounts for the important
interaction current in the $NN$ bremsstrahlung reaction in a manner
that is consistent with the generalized Ward--Takahashi identity
(WTI), which ensures gauge invariance at the microscopic level. This
feature is absent from all earlier models.

%
\begin{figure*}[t!]\centering
\includegraphics[width=.9\textwidth,clip=]{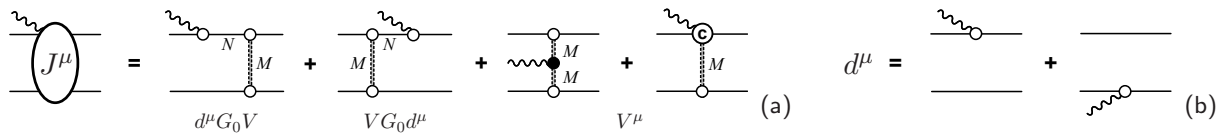}
\caption{\label{fig:model}%
(a) Basic photon production current $J^\mu$ used to describe the
$pp\to pp\gamma$ reaction in the present work; analogous photon
couplings along the lower nucleon line are omitted. $N$ denotes an
intermediate nucleon, and $M$ incorporates all the meson exchanges.
As indicated by the symbols below the diagrams, the first two
describe the nucleonic current, while the meson-exchange current is
depicted by the third. The fourth diagram contains the $N M\to N
\gamma$ four-point contact current which, together with the
meson-exchange current, constitutes the interaction current $V^\mu$
of Eq.~(\ref{eq:1J}). Diagram (b) shows the photon coupling to both
intermediate nucleons as subsumed in the dual current $d^\mu$.}
\end{figure*}
%
%

Following Ref.~\cite{HN10}, one starts from the nucleon-nucleon
$T$-matrix determined by the relativistic Bethe--Salpeter (BS)
equation
\begin{equation}
T = V + VG_0T = V + TG_0V~,
\label{eq:BbS}
\end{equation}
where $V$ represents the driving two-nucleon potential. The
two-nucleon propagator, $G_0=S_1 S_2$, describes the intermediate
propagation of two free non-interacting nucleons (with individual
Feynman propagators $S_i$, $i=1,2$) sharing the given fixed reaction
energy. This relativistic four-dimensional equation is then reduced
in a covariant manner to the three-dimensional Blankenbecler--Sugar
(BbS) equation~\cite{BbS,Bonn_NN} by replacing the propagator $G_0$
by   $G_0 \to g_0$ where $g_0$ restricts the intermediate two
nucleons to be on their mass-shells in a manner that preserves the
(relativistic) unitarity of the equation.

The driving potential $V$ used here is based on the
one-boson-exchange model developed by the Bonn group~\cite{Bonn_NN},
which contains nucleonic and mesonic degrees of freedom. In addition
to reproducing the low-energy $pp$ scattering data and the deuteron
properties, the resulting $NN$ interaction fits the $NN$ phase-shifts
up to the threshold for pion production. This version is used, rather
than a more modern potential, because the necessary interaction
current that is fully consistent with this potential is already
available from the work of Ref.~\cite{NH09}, where it was shown to be
crucial in resolving longstanding theoretical issues with the KVI
data~\cite{KVI02}.

By coupling the photon to the system of two interacting nucleons, it
can be shown, again following Ref.~\cite{HN10}, that the resulting
bremsstrahlung amplitude may be written as
\begin{equation}
M^\mu = (Tg_0+1)J^\mu(1+g_0T)~,
\label{eq:1}
\end{equation}
where the final-state (FSI) and initial-state interactions (ISI) are
included through the $NN$ $T$-matrices on the left and right,
respectively.

The basic photon production current from the two nucleons
\begin{equation}
J^\mu = d^\mu G_0 V +VG_0 d^\mu +V^\mu
\label{eq:1J}
\end{equation}
contains nucleonic and mesonic terms as well as a four-point contact-type term,
as illustrated in Fig.~\ref{fig:model}(a).

The two disconnected nucleonic terms, shown in
Fig.~\ref{fig:model}(b) and subsumed in the dual current $d^\mu$, are
given by
\begin{equation}
d^\mu \equiv  \Gamma_1^\mu (\delta_2 S_2^{-1}) + (\delta_1
S_1^{-1}) \Gamma_2^\mu\,. \label{OBodyC}
\end{equation}
Here $\Gamma_i^\mu$ is the $NN\gamma$ vertex for nucleon $i\ (=1,2)$,
$S_i$ denotes the propagator of the nucleon $i$, and $\delta_i$
represents an implied $\delta$-function that ensures that the
incoming and outgoing momenta of the intermediate nucleon $i$ are
identical.

We mention that the dynamical structure of this formulation takes
care of the fact that the translation of the three-dimensional BbS
reduction to the bremsstrahlung reaction must be implemented such
that a physical photon cannot couple to a nucleon that is on-shell
before \textit{and} after the coupling takes place. For more details
regarding this non-trivial issue, see Ref.~\cite{HN10}.

The $V^\mu$ of Eq.~\eqref{eq:1J} describes the photon coupling
to the internal mechanisms of the interaction $V$, i.e., it
corresponds to the interaction current. For a
one-boson-exchange model of the $NN$ interaction, such as that
employed here, $V^\mu$ consists of mesonic and four-point
contact currents. Unlike the case of proton-neutron
bremsstrahlung, where there is a large mesonic current
contribution~\cite{BF73,N89}, this is to a large extent
suppressed for proton-proton bremsstrahlung because only
neutral mesons can then be exchanged. The dominant mesonic
current contributions that we include arise from the anomalous
$v\pi\gamma$ couplings ($v=\rho, \omega$). These transitions
are transverse and thus cannot be obtained by simply coupling
the photon to the underlying $NN$ $T$-matrix; they must be
inserted by hand into $J^\mu$.

The four-point contact current appears as a consequence of imposing
gauge invariance in the form of the generalized WTI on the resulting
amplitude. Note that the $v\pi\gamma$ meson-exchange currents have no
influence on this because they are purely transverse. In general,
contact-type currents have very complicated microscopic dynamical
structures~\cite{H97} that cannot be taken into account explicitly at
present. Instead, one must revert to employing generalized
\textit{phenomenological} contact currents, constructed such that the
full reaction amplitude satisfies the generalized WTI, which is
necessary to ensure full gauge invariance at the microscopic level.
As a consequence, no unique determination of the reaction amplitude
is possible since the WTI does not constrain the transverse part of
the amplitude. In the present case, the dynamics of the hadron
interactions is described in terms of \textit{phenomenological} form
factors. The resulting phenomenological four-point interaction
currents that describe the interaction of the photon with this
hadronic three-point function, therefore, are constructed purely in
terms of these hadron form factors~\cite{HN10,NH09}.

In this paper we use our dynamical model in the analysis of the $pp
\to pp\gamma$ reaction data at a proton incident energy of 310~MeV.
Although this is well below the maximum of $\Delta$ production at
about 650~MeV, earlier analyses~\cite{dJNL95} (see also
Ref.~\cite{HN10}) of the TRIUMF data at 280~MeV~\cite{MIC1990} show
that its inclusion can improve the agreement with data in certain
geometries. We therefore investigate the effect of introducing the
$\Delta$ in a minimal fashion, following the application section of
Ref.~\cite{HN10}, by implementing the $\Delta$ contributions in
$J^\mu$ at the tree-level. For this purpose two more terms, analogous
to the first two on the r.h.s.\ of Fig.~\ref{fig:model}(a), are
added, with the $\Delta$ resonance replacing the intermediate nucleon
$N$. This $\Delta$ resonance current has no bearing on gauge
invariance because the $\Delta N\gamma$ transition vertex is purely
transverse. However, in a full $NN\rightleftharpoons\Delta N$
coupled-channels approach, in addition to the tree-level $\Delta$
resonance current considered here, there will also be additional
box-type contributions with intermediate $\Delta N$ and $\Delta
\Delta$ pairs that produce purely transverse five-point contact-type
contributions to the interaction current $V^\mu$~\cite{HN10}. At this
stage, therefore, the present minimal tree-level inclusion of the
$\Delta$ currents should be considered only exploratory.

The Bonn potential employed in the present study for generating the
nucleon-nucleon $T$-matrix is given in momentum space; it is
therefore non-trivial to include the Coulomb interaction. Coulomb
effects have been investigated in $pp$ bremsstrahlung in the
past~\cite{HNSA95}. However, the associated distortions are mainly
relevant at very small $pp$ invariant masses, a regime which has not
been well sampled in most of the earlier experiments.

In order to test the influence of the Coulomb interaction over the
wider acceptance of the present experiment, we also consider the $NN$
interaction based on the Paris potential~\cite{Paris_NN}. The Paris
work was carried out in coordinate space and includes fully the
Coulomb interaction~\cite{HNSA95} but only within the framework of
the non-relativistic Lippman--Schwinger equation. We have therefore
formally transformed this into the relativistic BbS equation by a
proper redefinition of the potential, through the so-called minimal
relativity factor~\cite{HN92}, in order to be able to use this
interaction consistently within the present relativistic approach.
The transformed interaction reproduces the same nucleon-nucleon
observables for relativistic kinematics as the original one for
non-relativistic kinematics.

One shortcoming in the present approach for incorporating the Paris
potential is that, for simplicity, we have retained the production
current $J^\mu$ calculated from the Bonn potential. As a result, the
consistency of the initial and final state interactions with the
production current $J^\mu$ is lost but, for the purpose of checking
the Coulomb effects, this inconsistency is not of major concern.

%
%
\section{Experiment}
\label{sec:experiment}%

The data of the present experiment were obtained at the The Svedberg
Laboratory in Uppsala, where a 48~MeV proton beam from the cyclotron
was injected into the CELSIUS ring~\cite{EKS88}, accelerated to
310~MeV and then stored. An average beam-on intensity of 3~mA was
achieved during an experimental data-taking period of approximately
100 hours.

The measurements were carried out at the PROMICE-WASA
facility~\cite{CAL96} and results at small proton-proton excitation
energies have already been published~\cite{JOH2009}. Furthermore, the
$pp \to pp\gamma$ data were obtained simultaneously with those on $pp
\to pp\pi^0$~\cite{ZLO98}, whose results were reported in greater
detail in Ref.~\cite{BIL01}. Since the detector assembly and the
measurement techniques were identical in the two experiments, and the
experimental procedures and data analysis differed only in minor
details, the description will here be kept quite brief.

An internal gas-jet hydrogen target, with a density of about $2\times
10^{14}$~cm$^{-2}$, was used in conjunction with the stored proton
beam. By operating an electron cooler throughout the experiment, the
background was reduced significantly and the counting rate increased
through an improvement in the beam-target overlap.

Even though the PROMICE-WASA facility was equipped to detect high
energy photons, in the bremsstrahlung study reported here, only
protons were measured in the final state. After exiting the
scattering chamber, the protons passed through a forward window
counter (FWC), a tracker, a forward trigger hodoscope (FTH) and
usually stopped in a forward range hodoscope (FRH). The four-quadrant
scintillator of the FWC eliminated most of the beam-halo background.
In order to be accepted by the main trigger, coincident protons must
appear in different quadrants. Events with protons in the same
quadrant were allowed by a secondary trigger but, in accord with
Monte Carlo expectations, these were very few in number and were not
considered in the subsequent analysis.

Information on the proton angles was extracted from the FTH and, even
more precisely, from the tracker. Events with polar angles between
about $3^\circ$ and $22^\circ$ were recorded. As described fully in
Ref.~\cite{BIL01}, the energy associated with a proton track was
deduced from a combination of the calculated angle-dependent range up
to the entrance of the stopping scintillator and the measured light
output of that detector. A few protons stopped in one of the thin
dead regions between the scintillator planes and these were then
assigned the energy corresponding to the midpoint of the dead layer.

As an extra check on the particle identification, it was further
required that both protons of an accepted event penetrate at least
into the second layer of the FTH, which consists of 24 spiral
scintillator segments. The minimum energy of each proton was
therefore 38~MeV. This condition meant that all coincident pairs of
protons stopped in the second FRH scintillator or earlier so that
there was effectively no high-energy limitation imposed by the design
of the apparatus.

\begin{figure}[t!]\centering
\includegraphics[width=.99\columnwidth,clip=]{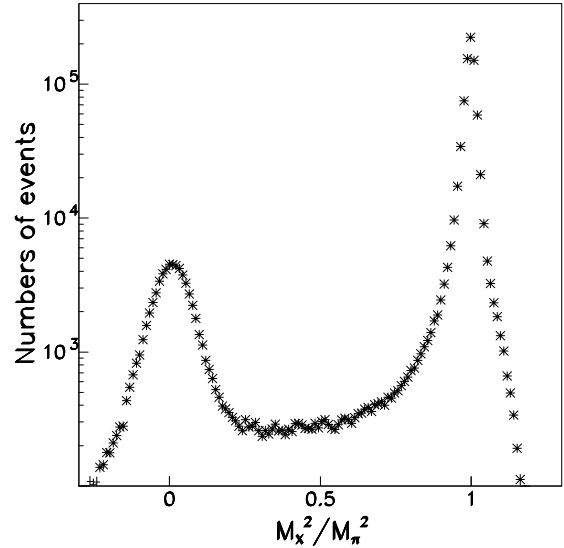}
\caption{\label{fig:mmsq} Distribution in the square of the missing
mass in the $pp\to ppX$ reaction presented in units of the neutral
pion mass. Clear peaks are seen, arising from the $pp\to pp\pi^0$ and
$pp\to pp\gamma$ reactions, sitting on a slowly varying background.}
\end{figure}

In the missing-mass distribution of the $pp \to ppX^0$ reaction shown in
Fig.~\ref{fig:mmsq}, there are two clear peaks corresponding to $X^0=\gamma$
and $X^0=\pi^0$, with very little overlap. Before making any detailed cuts,
these peaks contained in total 66,521 $pp\gamma$ and 861,449 $pp\pi^0$
candidates. The exclusion of events affected by the detector gaps, and those
where the proton time difference fell outside a 65~ns band, eliminated 7.3\%
and 1.5\% of these, respectively. There is only a small ($\approx 5\%$)
background under the $\gamma$ peak that arises mainly from the rescattering of
one of the protons from a pion-production reaction. The maximum polar angle of
protons from $\pi^0$ production depends sensitively upon the proton beam energy
$T_p$. A measurement of this angle, which was close to $18^\circ$, showed that
$T_p=309.7\pm0.3$~MeV.

The width of the $\gamma$ peak is $\sigma(M_X^2) =
0.056M_{\pi^0}^2$. By retaining only events at a little over
the two FWHM level, namely $|M_X^2/M_{\pi^0}^2|<0.137$, to a
good approximation this cut compensates for the neglect of the
small background contribution~\cite{JOH2009}.

\begin{figure}[t!]\centering
\includegraphics[width=.99\columnwidth,clip=]{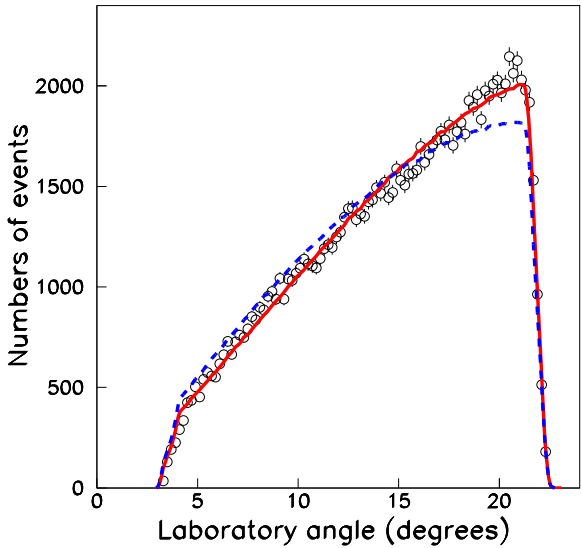}
\caption{\label{fig:angle_fit} (Color online) Angular
distribution of the final protons from the $pp\to pp\gamma$
reaction in the laboratory (open circles) compared to Monte
Carlo simulations based upon the HN model~\cite{HN10}, using
the Bonn potential as input (red solid curve). Similar
predictions obtained with a phase-space model (blue dashed
curve) are also shown. Both sets of predictions were normalized
to the total number of events.}
\end{figure}

The angular distribution of the protons in the laboratory system for the
selected $pp\to pp\gamma$ events is shown in Fig.~\ref{fig:angle_fit}. In spite
of a slight but significant misalignment between the beam and the detector
axes, the angular cutoffs at both small and large angles are quite sharp and
very well reproduced by the Monte Carlo simulation that used the Bonn $pp$
potential in the model described in Sec.~\ref{sec:model}. The phase-space
simulation gives a marginally poorer representation, especially at large
angles. In both cases the predictions have been normalized to the total number
of events.

\begin{figure}[t!]\centering
\includegraphics[width=.99\columnwidth,clip=]{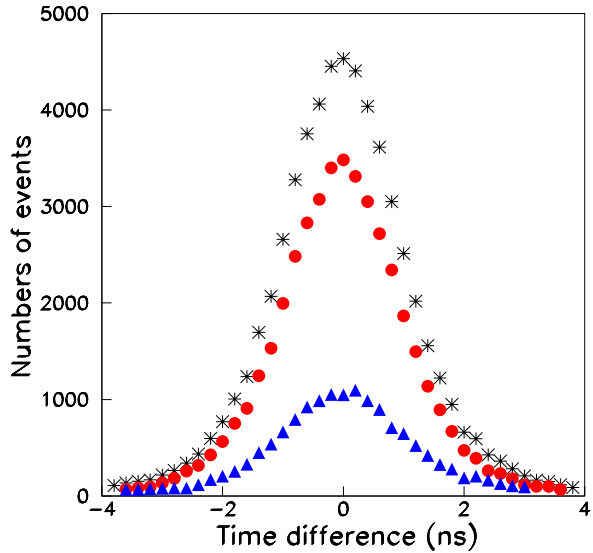}
\caption{\label{fig:timing} (Color online) Time difference between
the two protons emerging from the target. The peaks have similar
widths for photons in the forward (blue triangles) and backward (red
circles) CM hemisphere, as well as their sum (black crosses). Events
in the interval $\pm 1.8$~ns were retained for the subsequent
analysis.}
\end{figure}

For each of the emerging protons, a timing signal was extracted from the first
of the FTH spiral detectors. The time at the target position was then estimated
using the information on the particle energy, the hit position in the
scintillators, and the time-of-flight. The calibration, which was improved over
that used in Ref.~\cite{BIL01}, led to a distribution for the time difference
between the two protons with a peak width of 1.1~ns FWHM. This was essentially
the same for both the forward- and backward-going photons, though a correction
was introduced to compensate for a slight offset of 0.2~ns in the forward case.
Cuts at $\pm 1.8$~ns applied to the data of Fig.~\ref{fig:timing} reduced the
number of accidental coincidences to less than 1\% so that it was then
justified to employ a kinematic fitting. This was achieved by adjusting the
energies of the two protons to give zero missing mass. Taken together with the
sharper time difference cut, this reduced the number of candidates to the
58,026 $pp\gamma$ events that were used in the subsequent analysis.

%
%
\section{Analysis}
\label{sec:analysis}%

In the case of a production reaction like $pp\to pp\gamma$, the
unpolarized cross section is a function of four independent
variables. The standard set chosen for the analysis consists of
\begin{itemize}
\item $E_{pp}$: the excitation energy in the final $pp$
    system, %
\item $\theta_{\gamma}$: the CM production angle of the
    photon, %
\item $\theta_q$: the CM polar angle of the $pp$ relative
    momentum $\vec{q}$ with respect to the beam direction, %
\item $\varphi_q$: the azimuthal angle between $\vec{q}$
    and the photon momentum.
\end{itemize}
Other variables, such as the laboratory proton angle that was used in
the construction of Fig.~\ref{fig:angle_fit}, can be expressed in
terms of these four quantities.

\begin{figure}[t!]\centering
\includegraphics[width=.99\columnwidth,clip=]{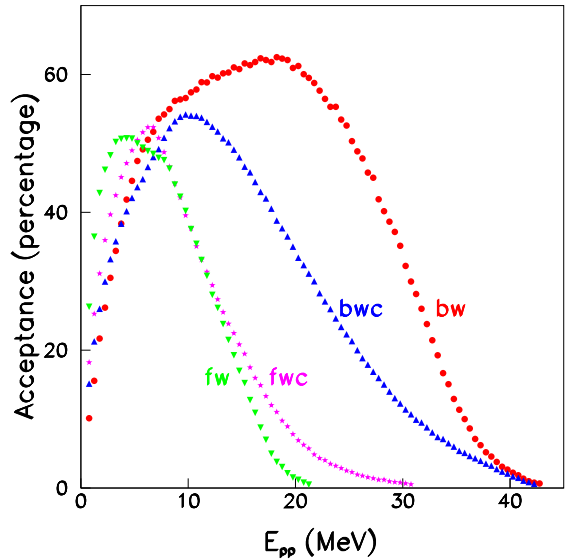}
\caption{\label{fig:acceptance} (Color online) Monte Carlo estimates
of the percentage acceptance for $pp\to pp\gamma$ at $T_p=310$~MeV
obtained using the HN model~\cite{HN10} with the Bonn potential. The
results are divided into four regions in the photon CM angle
$\theta_{\gamma}$ defined in the text, namely the backwards (red
circles), the backwards central (blue triangles), the forwards
central (magenta stars), and the forwards (green inverted
triangles).}
\end{figure}

In order to convert the observed numbers of events into cross
sections, knowledge of the detector acceptance is needed in the
four--dimensional space. This was achieved using a Monte Carlo
simulation, where the detector system was described in great
geometric detail. Identical cuts were then placed on the simulated
and experimental events. Frequently only phase-space was used in the
simulations but, in principle, the acceptances might depend
significantly on the actual reaction probability. Estimates were
therefore made, not only for simple phase space, but also for a
realistic reaction matrix, assumed to be represented by the HN
model~\cite{HN10}. In the latter case the program interpolated within
a lookup table of the reaction matrix yielded by this model in the
four standard variables. Values of the acceptance in the
($E_{pp},\,\theta_{\gamma}$) space are shown in
Fig.~\ref{fig:acceptance}.

It should first be noted that the PROMICE--WASA detector only
registered protons with laboratory angles less than $22^{\circ}$ and
at 310~MeV this means that only the region $E_{pp}<42$~MeV was
sampled. Although at low $E_{pp}$ the acceptance could be quite
large, being up to 60\%, this decreased to much lower values at
higher $E_{pp}$ and small $\theta_{\gamma}$. This is illustrated in
Fig.~\ref{fig:acceptance} by showing the acceptance as a function of
$E_{pp}$ in four ranges of the photon (CM) angle. These are,
respectively, the backwards $-1<\cos\theta_{\gamma}<-0.8$ (bw), the
backwards central $-0.8<\cos\theta_{\gamma}<0$ (bwc), the forwards
central $0<\cos\theta_{\gamma}<0.8$ (fwc), and the forwards
$0.8<\cos\theta_{\gamma}<1$ (fw) regions. For photons emitted in the
forward hemisphere, the recoiling protons are slower and a greater
fraction emerge at larger angles than allowed for in design of the
PROMICE--WASA detector, and this leads to a more severe cut at high
$E_{pp}$. Protons from these events are also more likely to be
distorted by secondary interactions. On the other hand, it also means
that the beam--pipe effect kicks in at lower $E_{pp}$, which is also
clearly seen in Fig.~\ref{fig:acceptance}.

\begin{figure}[t!]\centering
\includegraphics[width=.99\columnwidth,clip=]{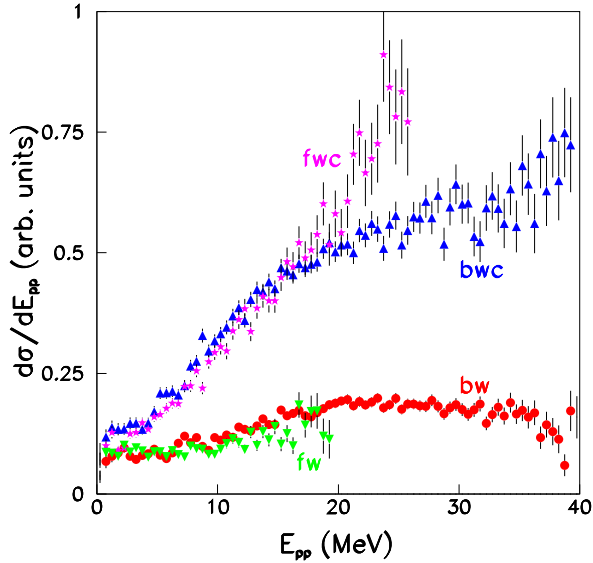}
\caption{\label{fig:fwbw} (Color online) Cross section for $pp\to
pp\gamma$ at $T_p=310$~MeV in arbitrary units, obtained using the HN
model~\cite{HN10} with the Bonn potential to evaluate the acceptance.
The experimental data are divided into the same four angular regions
as in Fig.~\ref{fig:acceptance}. Data were only plotted when the
acceptance was estimated to be above 2\%.}
\end{figure}

Due to the identical nature of the protons in the entrance channel,
the $pp\to pp\gamma$ cross section is symmetric in the CM system
around $\theta_{\gamma}=90^\circ$. The effects of the variation of
the acceptance with $\theta_\gamma$ at large $E_{pp}$ can also be
seen in Fig.~\ref{fig:fwbw}, which shows the cross sections extracted
as functions of $E_{pp}$ in the same four regions of
$\cos\theta_\gamma$ used in Fig.~\ref{fig:acceptance}. In all cases
the data were terminated when the estimated acceptance dropped below
2\%, and this occurred much earlier for small values of
$\theta_\gamma$. Except at the edges of the acceptance, the cross
sections deduced using a phase space model to evaluate the acceptance
differed only marginally from those obtained on the basis of the
dynamical model.

The crucial forward/backward symmetry is clearly respected to within
the uncertainties for $E_{pp}<20$~MeV but between 22 and 26~MeV there
is some deviation, which is more apparent in the angular
distributions to be presented in Sec.~\ref{sec:results}. On general
grounds one would expect the data from the forward photon hemisphere
to be less reliable because the associated protons are less energetic
and can emerge at larger angles. The statistics in the backward
hemisphere are also much larger.

The integrated luminosity of 340~nb$^{-1}$ was derived from a
comparison of elastic proton--proton scattering results measured in
parallel with tabulated cross sections, as described in
Ref.~\cite{BIL01}. Due to the large pre--scaling factor associated
with the $pp$ trigger used, an error bar of about 10\% must be
associated with this value. This includes also effects connected with
the evaluation of the proton acceptance in the apparatus and any
uncertainty in the $pp$ database used in the comparison.

Of the other systematic uncertainties discussed in
Ref.~\cite{BIL01}, proton rescattering in the detector material
might contribute 2\%, as might the treatment of the background
under the $\gamma$ peak. Although the PROMICE--WASA geometric
acceptance is very good, the extrapolation to unexplored
regions and its dependence upon reaction models can give up to
3\%, though this depends upon the value of $E_{pp}$. The known
systematic uncertainty is therefore judged to be $\approx 15\%$
overall. However, despite the care taken with the calibrations
and the evaluation of the acceptance, the forward/backward
symmetry is not completely respected at high $E_{pp}$, as
evidenced by the divergence between the fwc and bwc data in
Fig.~\ref{fig:fwbw}. We therefore cannot exclude larger
systematic uncertainties even in the backward photon hemisphere
for $E_{pp}\gtrsim 30$~MeV.

%
%
\section{Results}
\label{sec:results}%

Over 58,000 kinematically well-defined $pp\to pp\gamma$ events
are available for analysis in terms of the four-dimensional
differential cross section, as described in
Sec.~\ref{sec:analysis}. In the present paper only one single
differential and three double differential distributions are
presented. Data points are shown if the acceptance at this
point is estimated to be larger than 2\%. Only the statistical
uncertainties are shown explicitly by error bars and these do
not include the $\approx 15\%$ overall systematic effects. The
azimuthal dependence of the data can be quite strong but, since
this seems mainly to be a reflection of the acceptance, it is
not further investigated. The data would allow explorations of
higher dimensionality, but further guidance from theory would
be necessary to exploit this fruitfully.

\begin{figure*}[t!]\centering
\includegraphics[width=.9\textwidth,clip=]{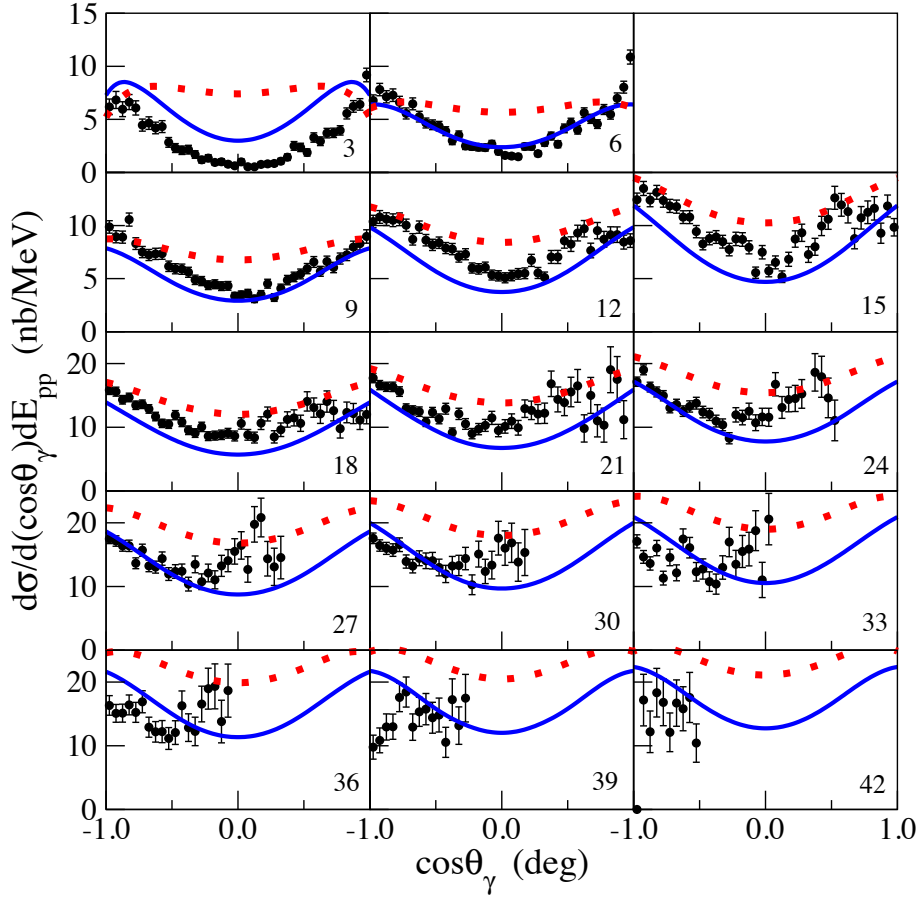}
\caption{\label{fig:ang1} (Color online) Differential cross section
for $pp\to pp\gamma$ at 310~MeV. The data have been averaged over
3~MeV bins in $E_{pp}$ and the upper end of each interval is
indicated by the number at the bottom right of the corresponding
panel. The curves correspond to the absolute predictions of the
Haberzettl and Nakayama model~\cite{HN10} of Sec.~\ref{sec:model},
using the Bonn potential~\cite{Bonn_NN}. The solid (blue) one
represents the model without any $\Delta$ contribution whereas for
the dashed (red) one the $\Delta$ has been switched on. The latter
has been included in a minimal fashion, as explained in
Sec.~\ref{sec:model}. }
\end{figure*}

The center--of--mass differential cross section in the photon
angle is presented in Fig.~\ref{fig:ang1} averaged over 3~MeV
bins in the $pp$ excitation energy from 0\,--\,3~MeV to
39\,--\,42~MeV, with the upper end of each interval being
indicated in the relevant panel. The angular cuts on the data
clearly reflect the acceptance dependence presented in
Fig.~\ref{fig:acceptance}.

As already reported~\cite{JOH2009}, the data for $E_{pp}<3$~MeV show
a strong minimum at $\theta_\gamma=90^\circ$ and an almost pure
$\cos^2\theta_\gamma$ behavior. The level of this minimum rises as
$E_{pp}$ increases. The data show that the $\cos^4\theta_\gamma$ term
is generally small and its strength cannot be determined with
precision.

The full dynamical model of Haberzettl and Nakayama~\cite{HN10}
of Sec.~\ref{sec:model} has been evaluated using the Bonn
potential~\cite{Bonn_NN} without the $\Delta$ contribution. The
curves are consistent with the shapes of the angular
distributions as measured in the backward hemisphere for all
except the lowest $E_{pp}$ bin. The strengths are also well
described, especially in view of the uncertainties in the
absolute scales of both the theory and experiment. A minimum is
predicted at $90^\circ$ for all energy bins but for
$E_{pp}<3$~MeV this is not sufficiently deep and there seems to
be no sign there of the leveling off near the forward/backward
directions expected from the theoretical model.

The theoretical predictions are, of course, sensitive to the
assumptions in the model. Thus, when the ``minimal'' inclusion of the
$\Delta$ contribution is switched on in the calculation, the effects
are surprisingly large and the agreement with the data is much
poorer, especially at low $E_{pp}$ where the central minimum is
largely absent. The shapes are far less changed at high $E_{pp}$ and
the data there can be well reproduced if the predictions are scaled
by a factor of $\approx 0.8$.

The description of the angular distributions with the Paris
potential~\cite{Paris_NN} is very similar to that obtained with the
Bonn potential. The Coulomb effects that are included here are only
significant for very low $E_{pp}$ but this is also the region where
the theoretical model is least satisfactory.

\begin{figure*}[t!]\centering
\includegraphics[width=.999\textwidth,clip=]{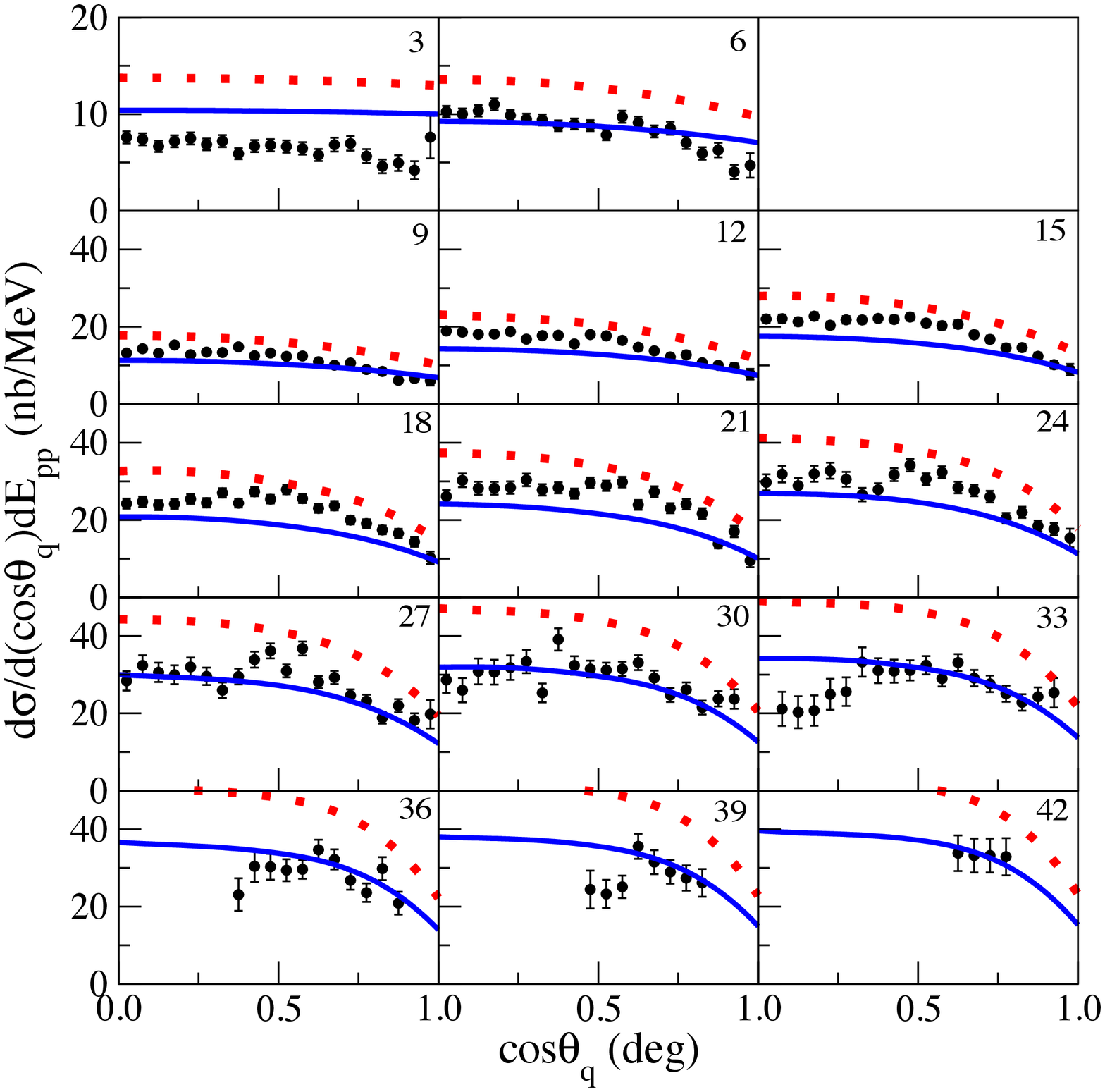}
\caption{\label{fig:ang2} (Color online) The distribution in the
angle between the $pp$ relative momentum vector $\vec{q}$ and that of
the incident beam direction in the overall CM frame. It should be
noted that there may be significant systematic measurement
uncertainties at low $E_{pp}$. Since the final protons are identical,
data are only shown in one hemisphere and the total cross section is
obtained by summing over this region. The curves correspond to the
predictions of the Haberzettl and Nakayama model~\cite{HN10} of
Sec.~\ref{sec:model} obtained using the Bonn
potential~\cite{Bonn_NN}. The notation is the same as in
Fig.~\ref{fig:ang1}. }
\end{figure*}

The second angular distribution to be discussed is that of the $pp$
relative momentum vector $\vec{q}$ with respect to that of the proton
beam in the overall CM frame. This is shown in Fig.~\ref{fig:ang2} in
the same 3~MeV bins that were used in Fig.~\ref{fig:ang1}. It is
difficult to measure the angles of the vector $\vec{q}$ when its
magnitude is small so that any apparent deviation from isotropy for
$E_{pp}<3$~MeV may not be significant.

The data for $E_{pp}>3$~MeV show clear evidence of a forward
dip and the predictions of the HN model~\cite{HN10} on the
basis of the Bonn potential follow this trend quantitatively.
As can be seen from Fig.~\ref{fig:ang2}, in this case there is
very little difference in the shape of the predictions whether
the $\Delta$ is included or not. The lack of data for small
$|\cos\theta_q|$ at high $E_{pp}$ is a consequence of the
limited acceptance in these regions.

\begin{figure*}[t!]\centering
\includegraphics[width=.999\textwidth,clip=]{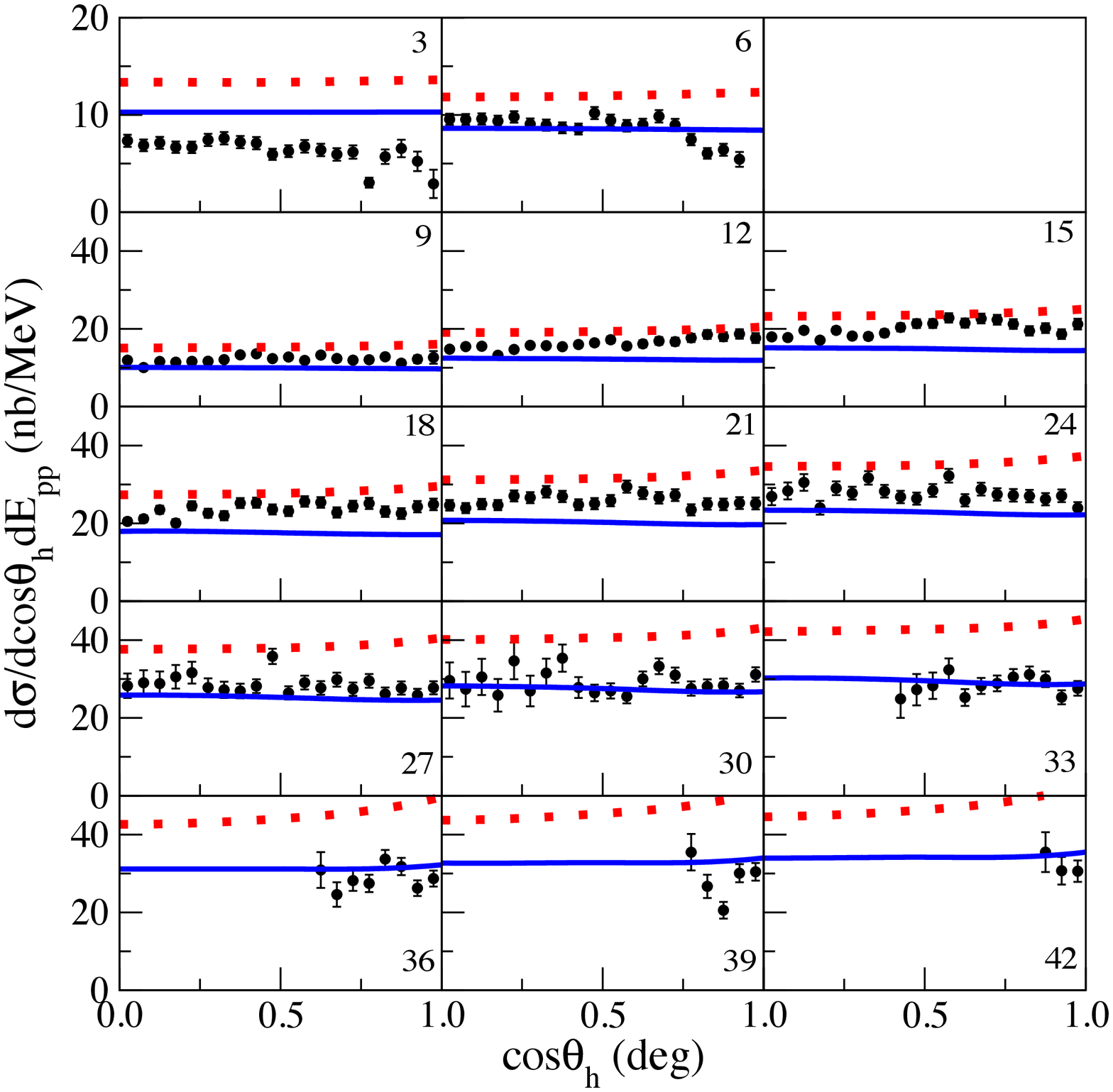}
\caption{\label{fig:ang3} (Color online) The distribution in the
helicity angle between the $pp$ relative momentum vector $\vec{q}$
and that of the photon in the frame of the recoiling $pp$ pair. The
notation is the same as in Fig.~\ref{fig:ang1}.}
\end{figure*}

The corresponding distribution of $\vec{q}$ with respect to the
photon direction is shown in Fig.~\ref{fig:ang3}. Although this is
evaluated in the frame of the recoiling $pp$ pair (the helicity
distribution), little change would be seen if this were replaced by
the overall center--of--mass frame. Both the data and the models
display fairly flat shapes, with the possible exception of the very
low energy bins, where the drop in the data for small helicity angle
$\theta_h$ may reflect the difficulty in measuring two protons when
they emerge with similar angles and momenta.

\begin{figure}[t!]\centering
\includegraphics[width=.95\columnwidth,clip=]{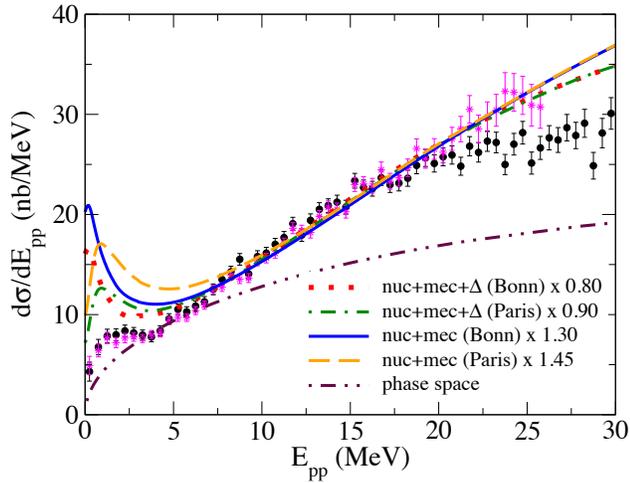}
\caption{\label{fig:fwbw2} (Color online)  Cross section for the
$pp\to pp\gamma$ reaction measured at $T_p=310$~MeV integrated over
all angles. The (black) circles were obtained by taking data from the
backward photon hemisphere and invoking the forward/backward
symmetry. For the (magenta) stars, data in the interval
$-1<\cos\theta_{\gamma}<0.8$ were used in the determination. Overall
systematic errors of $\pm 15\%$ are not shown. The solid (blue) curve
represents the evaluation of the model described in
Sec.~\ref{sec:model} using the Bonn $NN$ input without the inclusion
of any $\Delta$ contribution, whereas the red dots represent an
attempt to include this in a minimalist way. The (orange) dashed and
green chain curves are the analogous estimates based upon using the
Paris potential that includes the Coulomb repulsion. The curves are
scaled by the factors shown in the figure. The (maroon)
dot-dot-dashed curve illustrates the energy dependence to be expected
from a pure phase-space model. It is arbitrarily normalized at around
5~MeV, i.e., just above the region whose energy dependence is largely
driven by the strong $pp$ FSI.}
\end{figure}

Note that the data in the three angular distributions are different
representations of the same 58,026 events, so that the cross sections
integrated over angle must be identical. The ``holes'' seen at
various places of phase space in the diagrams indicate possible
sources of systematic errors. As a consequence one must conclude that
the integrated $pp\to pp\gamma$ cross section can only be safely
extracted when $E_{pp}\lesssim 30$~MeV. The energy variation of this
cross section is shown in Fig.~\ref{fig:fwbw2} as a function of
$E_{pp}$. For one of the sets of points, only results from the
backward photon hemisphere (bw + bwc) are used. The other set uses in
addition the fwc data and the forward/backward symmetry is only
invoked to derive the data in the very forward region.

The reasonable description of the angular distributions by the
theoretical model is translated into one of the integrated cross
section, where the predictions have been smeared over the $0.5$~MeV
bins used in the data presentation. In order to ensure agreement with
the data above 10~MeV, the theoretical results that included only the
nucleonic and meson-exchange current terms were scaled by factors of
1.30 and 1.45 for the Bonn and Paris evaluations, respectively. When
the $\Delta N$ intermediate states are included in the minimal way
described in Sec.~\ref{sec:model}, the corresponding factors are 0.80
and 0.90, respectively.

The drawback of using the Bonn potential is immediately apparent. The
unrealistic spike at very low $E_{pp}$ is significantly softened when
this is replaced by the Paris potential, which includes the Coulomb
repulsion in the $pp$ system. Otherwise there is little difference
between the predictions based upon the two potentials. In either case
the model seems to overestimate the cross section for the production
of the $^{1\!}S_0$ state of the two final protons. It must again be
stressed that in this region there can be delicate cancelations
amongst the contributions~\cite{LAG1989}. If the $^{1\!}S_0$
prediction were reduced, the energy dependence might be reproduced,
though with a slightly too low overall normalization.

The inclusion of the effects of the $\Delta$ in an approximate way
gives a rather similar description of the data in
Fig.~\ref{fig:fwbw2}. However, as mentioned already, the $\Delta$
effects beyond the tree-level that would arise in a consistent
$NN\rightleftharpoons\Delta N$ coupled channels approach~\cite{HN10},
have been ignored in the present calculations. Using the Paris rather
than the Bonn potential induces changes analogous to those seen in
the non-$\Delta$ scenario.

Phase space does not provide an acceptable description of the
$E_{pp}$ dependence of the integrated cross section shown in
Fig.~\ref{fig:fwbw2}. This is by no means unexpected because the
spin-parity constraints associated with the $^{1\!}S_0$ final state
are not built into such a naive approach.

The total $pp\to pp\gamma$ cross section integrated up to
$E_{pp}=30$~MeV is $\sigma(30) = (0.59\pm 0.09)~\mu$b, where the
statistical error is negligible and the quoted uncertainty is purely
systematic. In the COSY-TOF measurement at 293~MeV~\cite{BIL1998}
there were very few events with $E_{pp}>60$~MeV and a total cross
section of $\sigma=(3.5\pm0.3\pm0.7)~\mu$b was obtained by
extrapolating to the kinematic limit on the basis of a phase-space
variation. If, instead, the total cross section below 30~MeV is
estimated, a value of $\sigma(30) = (0.54\pm 0.12)~\mu$b is found.
The agreement between the COSY-TOF result and ours lies well within
the error bars.

\begin{figure}[t!]\centering
\includegraphics[width=.99\columnwidth,clip=]{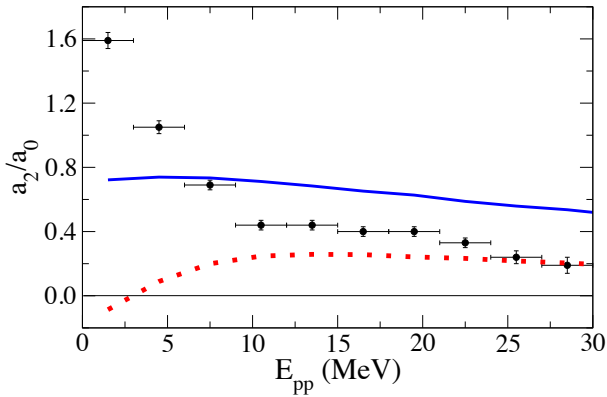}
\caption{\label{fig:fwbw4} (Color online) Ratio of the Legendre
coefficients defined by Eq.~\eqref{Leg}. Values extracted from
experimental data are compared to the predictions of the Haberzettl
and Nakayama model~\cite{HN10} with and without the minimalist
$\Delta$ inclusion (red dotted and solid blue lines, respectively). }
\end{figure}

In order to illustrate the variation of the angular dependence of the
photon with excitation energy, the differential cross section data of
Fig.~\ref{fig:ang1} have been fitted by
\begin{equation}
\label{Leg}
\frac{d\sigma}{d\cos\theta_\gamma}=\frac{1}{2}\sum_{n=0}^{2}a_{2n}P_{2n}(\cos\theta_\gamma)\,.
\end{equation}
Figure~\ref{fig:fwbw4} shows the ratio $a_2/a_0$ as a function of
$E_{pp}$ compared to the predictions of the present calculation based
on the HN model~\cite{HN10}. Neither the prediction with or without
the $\Delta$ contribution can describe the rise in $a_2/a_0$ at low
$E_{pp}$ and the inclusion of the $\Delta$ is particularly
disappointing in this region.

%
%
\section{Conclusions}
\label{sec:conclusions}%

We have presented here detailed measurements of bremsstrahlung
production in proton-proton collisions at a beam energy of
310~MeV. The differences between the data extracted from the
forward and backward photon hemispheres increases for
excitation energies above about 20~MeV and, in such cases, the
backward data are more reliable because of the faster protons
and the much higher acceptance. Using the forward/backward
symmetry of the reaction, full acceptance was achieved up to an
$E_{pp}\approx 30$~MeV and some information obtained even close
to the kinematic limit of 42~MeV imposed by design of the
apparatus. The big advantage of this experiment compared to
others undertaken above the pion-production threshold is the
large acceptance coupled with good statistics. It is therefore
not surprising that the values extracted for the cross sections
depend very little whether one uses phase space or the HN model
to estimate the acceptance. However, to go to higher $E_{pp}$
one would need larger counters than those provided by the
PROMICE-WASA setup.

Away from the very small $E_{pp}$ region, the dynamical model of
Haberzettl and Nakayama~\cite{HN10}, whose main points are summarized
in Sec.~\ref{sec:model}, is rather successful in describing all the
experimental results as functions of $E_{pp}$ provided an overall
scaling factor close to unity is applied. As well as the integrated
cross section, these include the angular distributions of the photon
and those of the $pp$ relative momentum with respect to the beam
direction and to that of the recoil photon. It is very gratifying to
note that any rescaling of the predictions required to achieve this
success is well within the combined experimental and theoretical
uncertainties. The latter are clearly very hard to quantify but they
must at least encompass the differences between the inclusion or not
of the $\Delta$ contributions.

The situation at low $E_{pp}$ is more uncertain because there are
significant cancelations amongst the driving terms~\cite{LAG1989} and
the theoretical results are therefore much more sensitive to small
contributions. Using the Paris rather than the Bonn potential in the
evaluation of the model allows the Coulomb interaction to be included
and this does smooth the predictions slightly at low $E_{pp}$.
However, it must be noted that the switch from Bonn to Paris was not
done fully consistently.

The good agreement between theory and experiment was obtained without
considering any effects that might arise from the virtual excitation
of the $\Delta$ isobar. Although the basic model was tuned to
describe the 190~MeV KVI data~\cite{KVI02}, by 310~MeV the influence
of the $\Delta$ might start to be felt. In this context it should be
noted that the introduction of the $\Delta$ isobar~\cite{dJNL95,HN10}
seems to improve the theoretical description of the 280~MeV TRIUMF
data~\cite{MIC1990}. However, we find that the inclusion of the
$\Delta$ effects in the minimal way described in Sec.~\ref{sec:model}
actually makes the agreement with the shapes of the photon angular
distribution worse for $E_{pp}\lesssim 12$~MeV and this difference is
somewhat puzzling. We view it more as an indication that $\Delta$
effects are not very well understood and that minimal inclusion is
not warranted in these instances, rather than as a measure of
theoretical errors. Further theoretical work is clearly needed and
the results might be improved by introducing a phenomenological
five-point contact current to take account of the $\Delta N$ box
diagrams, as explained in Ref.~\cite{HN10}.

At higher $E_{pp}$ the differences are less important and the four
versions of the model give very similar integrated cross sections in
Fig.~\ref{fig:fwbw2} provided that they are scaled by factors that
are all fairly close to unity. With the scaling factors as shown, the
predictions start to differ from the more reliable backward-angle
data above 20~MeV. However, in view of the 15\% overall systematic
error in the data and the arbitrariness in the scaling factors one
cannot draw firm conclusions as to the significance of this.

The very small $E_{pp}$ region has been studied near the forward
direction with the COSY-ANKE spectrometer up to
800~MeV~\cite{KOM2008,TSI2010}. The results show an energy dependence
that suggests some influence from intermediate $\Delta N$ pairs.
However, for kinematic reasons, at these higher beam energies the
$\gamma$ and $\pi^0$ peaks, which are so prominent in the
missing-mass plot of Fig.~\ref{fig:mmsq}, merge and the extraction of
a $pp\to pp\gamma$ signal is much more delicate. Under these
conditions it may be necessary to measure the photon in coincidence
and data of this type from the COSY-WASA facility are currently being
analyzed at 500 and 550~MeV~\cite{ZLO2010}.

However, it should be stressed that data in the small $E_{pp}$ region
taken at well below the pion production threshold would also be very
valuable because the uncertainties regarding the inclusion of the
$\Delta$ contribution would then be minimized. Data with polarized
beam and target, along with results on the differential cross section
and analyzing power, would allow some of the electromagnetic
multipoles to be determined explicitly and therefore allow one to
identify defects in the models more clearly~\cite{NAK2010}.

The bremsstrahlung reaction is the simplest of all the
high-momentum-transfer reactions in proton-proton collisions at
intermediate energies. Unlike cases of meson production, there
is no need to consider the final state interaction of the meson
with one of the protons. The quality of the agreement between
the predictions of a modern bremsstrahlung model with the
high-statistics and high-acceptance data achieved at the
PROMICE-WASA facility is striking and should encourage further
experimental and theoretical work in the field.

%
%
\begin{acknowledgements}
We are very grateful to the TSL/ISV personnel and the PROMICE-WASA
collaboration for their support in this experiment. We acknowledge in
particular the great help provided by Jozef Z{\l}oma\'{n}czuk in the
early stages of this experiment and its analysis. We also thank the
Research Computer Center at the University of Georgia and the
Supercomputing Center at Forschungszentrum J\"ulich for providing
computing resources that were necessary to obtain the theoretical
results shown in the present paper. This work was supported by the
European Community under the ``Structuring the European Research
Area'' Specific Programme Research Infrastructures Action (Hadron
Physics, contact number RII3-cT-204-506078), the FFE grant No.\
41788390 (COSY-058), and by the Swedish Research Council.
\end{acknowledgements}

%
%

\end{document}